\documentclass{article}

\usepackage{arxiv}
\usepackage[utf8]{inputenc} 
\usepackage[T1]{fontenc}    
\usepackage{hyperref}       
\usepackage{url}            
\usepackage{booktabs}       
\usepackage{lipsum}
\usepackage{graphicx}
\usepackage{cite}
\usepackage{multirow}%
\usepackage{amsmath,amssymb,amsfonts}%
\usepackage{amsthm}%
\usepackage{mathrsfs}%
\usepackage[title]{appendix}%
\usepackage[dvipsnames]{xcolor}%
\usepackage{textcomp}%
\usepackage{booktabs}%
\usepackage{algorithm}%
\usepackage{algorithmicx}%
\usepackage{listings}%
\graphicspath{ {./images/} }
\usepackage{caption}
\usepackage{soul}
\usepackage{siunitx}
\usepackage{subcaption}
\usepackage{physics}
\usepackage{comment}

\AtBeginDocument{\RenewCommandCopy\qty\SI}
\usepackage{nomencl}
\makenomenclature

\title{Modeling considerations about a \\ microchannel heat sink}

\author{
  L. Chej \\
  Laboratory of Modeling and Computational Simulation\\
  Institute of Sciences, National University of General Sarmiento\\
  Los Polvorines, Buenos Aires, Argentina.\\
  National Scientific and Technical
  Research Council, Argentina.\\
  Corresponding author.\\
  lchej@campus.ungs.edu.ar \\
   \And
  A. Monastra, and M.F. Carusela \\
  Laboratory of Modeling and Computational Simulation\\
  Institute of Sciences, National University of General Sarmiento\\
  Los Polvorines, Buenos Aires, Argentina.\\
  National Scientific and Technical
  Research Council, Argentina.\\
}

\begin{document}
\maketitle

\begin{abstract}
Many computational studies on hotspot microfluidic cooling devices found in the literature rely on simplified assumptions and conventions that do not capture the full complexity of the conjugate thermal problem, such as constant thermophysical fluid properties, radiation and free air convection on the external walls. These assumptions are generally applied to typical microfluidic devices with a large number of microchannels and operating at Reynolds numbers between $100-1000$. A one microchannel chip is a suitable starting point to analyze more systematically the implications of these assumptions, in particular at lower Reynolds numbers. Although it is a simpler system, it has been studied experimentally and numerically as a basic block of a thermal microfluidic device.
In this work we analyze the modeling of the overall heat transfer from a hotspot to a microfluidic heat sink, focusing on the effect of the different thermal transfer mechanisms  (conduction, convection and radiation), temperature-dependent thermophysical properties of the fluid and the chip material.  
The study is developed as a function of the pressure difference applied to the system based on simulations performed using a Finite Volume Method.
Analyzing and comparing the different contributions to the energy losses, this work provides a critical discussion of the usually considered approximations, in order to make a reliable modeling of the overall thermal performance of a single rectangular straight channel embedded in a PDMS microfluidic chip.
\end{abstract}


\nomenclature{\(T_{\text{max}}\)}{Maximum temperature (K)}
\nomenclature{\(T_{\text{out}}\)}{Temperature at the outlet (K)}
\nomenclature{\(Re\)}{Reynolds number}
\nomenclature{\(\sigma\)}{Steffan-Boltzmann constant}
\nomenclature{\(\epsilon\)}{Emissivity ($W / (m^2 K^4)$) }
\nomenclature{\(h_{\text{rad}}\)}{Radiative heat transfer coefficient ($W / (m^2 K)$) }
\nomenclature{\(h_{\text{conv}}\)}{Convective heat transfer coefficient ($W / (m^2 K)$) }
\nomenclature{\(\text{PDMS}\)}{Polydimetilsiloxsane}
\nomenclature{\(G\)}{Constant Pressure Gradient (Pa/m)}
\nomenclature{\(V\)}{Volume Flow Rate ($m^3/s$)}
\nomenclature{\(\tau\)}{Critical Time ($s$)}
\nomenclature{\(\Dot{m}\)}{Mass Flow Rate ($kg/s$)}

\clearpage
\section{Introduction}\label{sec:introduction}
Microfluidic systems are ideal platforms for thermal management devices. 
Technological progress in the electronic field has produced a significant progress in the manufacture of more powerful devices with smaller dimensions \cite{barakoIntegratedNanomaterialsExtreme2018}.  However these achievements have the counterpart of demanding higher efficiencies in heat removal, due to the existence of larger heat fluxes (\SI{1000}{\watt/\centi\metre\squared}) with non-uniform distributions.  Multi-component electronic architecture technologies increasingly involve the presence of multiple localized hotspots for which conventional cooling methods are inadequate or insufficient, making thermal management a more complex and challenging issue \cite{yeExperimentalInvestigationsThermal2022,sohelmurshedCriticalReviewTraditional2017}.

In the literature can be found different proposals of thermal devices based on microgaps \cite{lorenzini_embedded_2016}, oblique fins \cite{lee_enhanced_2012}, sinusoidal-wavy microchannels \cite{khoshvaght-aliabadi_experimental_2016}, or microchannels with arrays of pillars \cite{grafner_flow_2022}, whose thermal transfer performances are characterized from direct or indirect measurements of flows and temperatures.  Usually the experimental studies are complemented and compared with computational simulations in order to improve and optimize the design of the devices \cite{carvalho_computational_2021,gao_fluid_2022,hung_optimal_2012}. This is a useful strategy that helps to avoid testing or prototyping which can be very costly processes. However, this is an effective methodology if most of the characteristics of the experimental setup and of the environmental conditions, usually complex, are taken into account in the computational modeling. 


In general, microfluidic devices are composed by a large number of microchannels with operative regimes characterized by Reynolds numbers in the range of 100 - 1000 \cite{sidik_overview_2017}. This is in order to overcome the pressure drops and to remove as much heat as possible by forced convection of the fluid at the outlet. 
The computational modeling of multichannel symmetric systems are in most cases firstly focused on a single microchannel unit and its immediate vicinity solid domain, in order to reduce the computational cost. This first approach gives a qualitative, and in many cases also quantitative, good estimation of the thermal transfer performance of the device by means of the response of a single microchannel unit \cite{kumar_numerical_2019, raghuraman_influence_2017}.

However, in the literature can also be found many theoretical, numerical and experimental studies concerning thermal transfer along one microchannel chip.
\cite{jungExperimentalInvestigationHeat2021, jungHeatTransferFlow2019, oudahExperimentalInvestigationEffect2020, wangExperimentalNumericalStudy2018, yiPDMSNanocompositesHeat2014}. The simplicity of this setup is sought for several reasons, among them to avoid the cost of manufacturing an entire device in a first stage, or to test the thermohydraulic performance in a faster and simpler way for different geometries or materials. In these particular cases, Re are now within the range of 1-100 \cite{jungExperimentalInvestigationHeat2021, jungHeatTransferFlow2019,yiPDMSNanocompositesHeat2014}. 

Considering these variations in the design and in the hydraulic operation range, certain assumptions that are widely accepted in most cases cannot be lightly addressed.
The abrupt high temperature gradient produced by the hotspot can lead to variations in the thermophysical properties of the fluid that, unlike in other cases, will not be negligible. We are interested in developing studies such as those carried out by \cite{liEffectsThermalProperty2007,kumarPhysicalEffectsVariable2018} in which the thermophysical properties of the fluid change with the local temperature, giving more accurate results with a non-significant additional computational cost.


To carry out a reliable analysis of the disagreements between measurements and simulations, it is necessary not only a qualitative but a quantitative critical approach to the considered approximations.  
A fundamental test to perform this critical analysis is to evaluate the energy losses due to the differnt heat transfer mechanisms present in the {\it whole} setup.  To illustrate this point, we considered a basic device constituted by a single microchannel with an embedded hotspot in a structure made of Polydimetilsiloxane (PDMS). We choose this material because it is widely used for the design of lab-on-chip microfluidic devices, including those for thermal management
\cite{yiPDMSNanocompositesHeat2014a,wuPolydimethylsiloxaneMicrofluidicChip2009,kumarThermalModelingDesign2013}, and also for its many benefits when compared to other feasible candidates such as glass or Si: low cost, ease of fabrication and biocompatibility \cite{linPDMSMicrofabricationDesign2021,joThreedimensionalMicrochannelFabrication2000}.  
Moreover, regarding the design of thermal devices, the high light transmittance of the PDMS, gives it a substantial advantage over Cu \cite{qu_analysis_2002} or Al \cite{ahmedOptimumThermalDesign2015} for making direct studies on the flow using Particle Image Velocimetry (PIV) or Laser Induced Fluorescence (LIF) \cite{jungHeatTransferFlow2019}. However, PDMS is a material with a low thermal conductivity. Therefore, its use in microfluidics heat exchangers for embedded electronics, and consequently embedded hotspots, requires an in-depth analysis of the effect on the overall thermal transfer properties of the {\it whole} device. 
The PDMS device proposed in this work, although simple, can be used to develop a critical analysis about the energy losses when the different thermal transfer mechanisms are considered as a function of the thermodynamic properties of the fluid  (e.g. compressibility, viscosity), the hydrodynamic conditions characterized by the Reynolds number, and the interplay with the environment. This analysis can be helpful when assumptions and approaches for the modeling of thermal transfer in microchannels are considered.

In this work we present a detailed discussion about some of these issues. In Section \ref{sec:physicalModel}, we show the computational models and the governing equations. In Subsection \ref{sec:solutionMethodology} we introduce the computational implementation of the models with a detailed discussion, and in Section \ref{sec:results} the results are presented.
Finally, in Section \ref{sec:concludingRemarks} some brief conclusions are presented.

\section{Physical model}\label{sec:physicalModel}

\begin{figure}[H]
     \centering
     \begin{subfigure}[b]{0.495\textwidth}
         \centering
         \includegraphics[width=\textwidth]{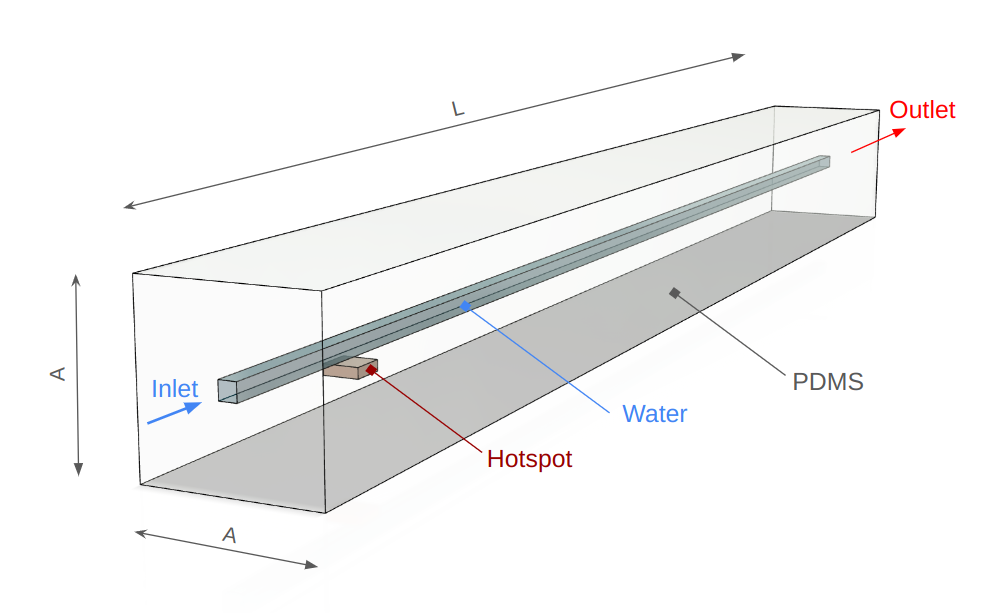}
         \caption{Schematic view of computational domain}
         \label{fig:schematic}
     \end{subfigure}
     \hfill
     \begin{subfigure}[b]{0.495\textwidth}
         \centering
         \includegraphics[width=\textwidth]{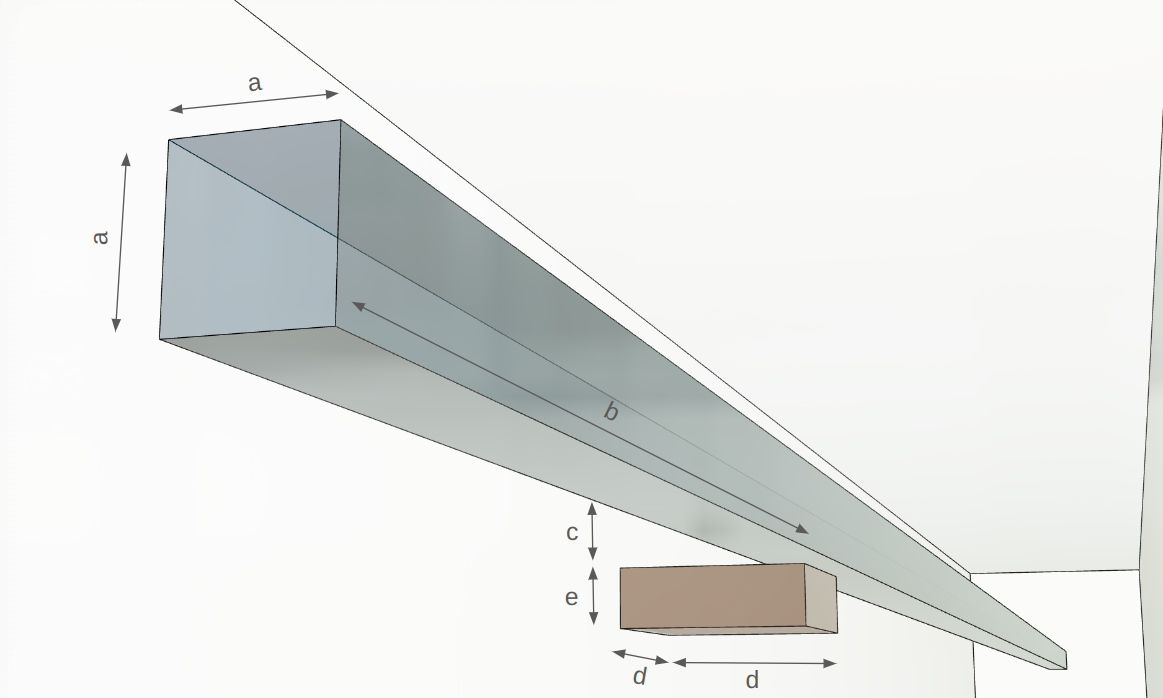}
         \caption{Schematic detail view}
         \label{fig:schematic detail}
     \end{subfigure}
        \caption{Gometric features}
        \label{fig:physicalModel}
\end{figure}

The structure of the proposed microfluidic device is illustrated in Fig.\ref{fig:schematic} and it consists of a microchannel and a hostspot embedded in a prism shaped-material made of PDMS whose dimensions are: \SI{5}{\milli\metre} $\times$ \SI{5}{\milli\metre} $\times$ \SI{50}{\milli\metre}. A square microchannel of \SI{500}{\micro\metre} side is located in the center of the PDMS structure as illustrated by Fig \ref{fig:schematic detail}. The coolant selected to circulate through the microchannel is water. It is considered a copper hotspot with dimensions $1$mm $\times$ $1$mm $\times$ \SI{300}{\micro\metre} that generates a heat power of \SI{0.2}{\watt}, typical values found in the literature. It is embedded in the PDMS at \SI{200}{\micro\metre} below the microchannel,  and placed at \SI{5}{\milli\metre} from the inlet.
\begin{table}[!ht]
\begin{center}
\caption{Model dimensions.\label{tab:dimensions}}
    \begin{tabular}{ccccccc}
    \hline
        A(mm) & L(mm) & a(\SI{}{\micro\metre}) & b(mm) & c(\SI{}{\micro\metre}) & d(mm) & e(\SI{}{\micro\metre}) \\ \hline
        5     & 50    & 500        & 5     & 200        & 1     & 300        \\ \hline
    \end{tabular}
\end{center}
\end{table}

To study the thermal transfer properties of the device, it is necessary to consider the set of conservation equations for the fluid and the material:

Conservation of mass
\begin{equation} \label{eq:mass}
    \pdv{\rho}{t}=\Vec{\nabla} \cdot (\rho \Vec{u})
\end{equation}
where $\rho$ is the mass density of the fluid and $\Vec{u}$ is the fluid velocity.

Conservation of momentum
\begin{equation} \label{eq:momentum}
    \rho \frac{D\Vec{u}}{Dt}=\Vec{\nabla} \cdot \overline{\overline{\sigma}} +\rho b
\end{equation}
$b$ represents the external forces per unit mass and $\overline{\overline{\sigma}}$ the stress tensor.

Conservation of energy
\begin{equation} \label{eq:energy}
    \rho \frac{De}{Dt}=-p \Vec{\nabla}\cdot\Vec{u}-\Vec{\nabla}\cdot\Vec{q}+\phi
\end{equation}
$e$ corresponds to the internal energy of the fluid per unit mass. The first term on the right hand side represents the work done by the external pressure $p$ on the fluid. The second term $\Vec{q}$ is the heat flux
\begin{equation}
    \Vec{q} = - \kappa_{f} \Vec{\nabla} T + \Vec{q}_{\text{rad}}
\end{equation}
with the first term due to heat conduction (Fourier's law) and the second due to electromagnetic radiation. $\kappa_{f}$ is the thermal conductivity of the coolant.
The last term
\begin{equation}
    \phi = \frac{1}{2}\mu \sum_{i,j}\left(\pdv{u_i}{x_j}+\pdv{u_j}{x_i}\right)^2 + \lambda (\Vec{\nabla}\cdot\Vec{u})^2
\end{equation}
is the dissipated kinetic energy. $\mu$ is the first (shear) viscosity and $\lambda$ is the second (bulk) viscosity.

For the material it is considered heat conduction given by Fourier's Law:
\begin{equation}
    \Vec{q} = -\kappa_{s} \Vec{\nabla} T+\Vec{q}_{\text{rad}}
\end{equation}
with $\kappa_{s}$ the thermal conductivity of the solid (PDMS).

In addition, temperature and heat flux must fulfill continuity at the solid-fluid interface:

\begin{equation}
    T_{\text{s}} = T_{\text{f}}
\end{equation}

\begin{equation}
    Q_{\text{s}} = -Q_{\text{f}}
\end{equation}
It is also assumed a non-slip boundary condition on the walls of the microchannel. 

The simpler approach is to analyze only the fluid in a microchannel. According to the Boussinesq formulation \cite{boussinesq1868} for a Poiseuille flow the volume flow rate and velocity profile can be calculated analytically. Assuming a rectangular cross section pipe with a height $0 \leq y \leq h$ and width $0 \leq x \leq w$, it is obtained: 
\begin{equation}\label{eq:rectangular}
\begin{aligned}
    u(x,y) = {} & \frac{G}{2\mu}y(h-y)\\
                & -\frac{4Gh^2}{\mu \pi^3}\sum_{n=1}^{\infty}\frac{1}{(2n-1)^3}\frac{\sinh(\beta_n x)+\sinh[\beta_n(w-x)]}{\sinh(\beta_n w)}\sin(\beta_n y)
\end{aligned}
\end{equation}
with $G = \frac{\Delta p}{L}$ as the constant pressure gradient in the flow direction and $\beta_n = \frac{(2n-1)\pi}{h}$. Then, the volume flow rate V can be defined as:
\begin{equation}
  V = \frac{Gh^3w}{12\mu}-\frac{16Gh^4}{\pi^5 \mu}\sum_{n=1}^{\infty}\frac{1}{(2n-1)^5} \frac{\cosh{  (\beta_nw)}-1}{sinh(\beta_nw)}
    \label{eq:caudal}
\end{equation}
As we consider a square cross section channel, $h=w$, from Eq.\ref{eq:caudal}, Reynolds number can be written as:
\begin{equation}
    Re = \frac{uD_h\rho}{\mu}= \frac{V\rho}{\mu w}
\end{equation}
where $D_{\text{h}}=\frac{A}{P}=w=h$ is the hydraulic diameter,  with $A$ is the square microchannel cross section area and $P$ is the channel wet perimeter.
Considering the characteristic parameters of our model, the Reynolds number takes values in the range $6-120$.

To address the heat transfer due to the presence of a  hotspot we make some general assumptions:
\begin{itemize}
    \item The fluid is Newtonian (shear stress proportional to velocity gradient)
    \item The gravity is neglected, so natural convection is not induced.
\end{itemize}
Given the length scales, pressure differences and viscosity, a laminar flow is induced in all cases. Within this framework it is possible to consider different models, going from simple idealized to more complex and realistic ones:
\begin{itemize}
    \item Model 1
\begin{itemize}
    \item Convective and radiative heat transfer from the external PDMS walls to the ambient are neglected.
    \item The fluid is incompressible and with constant thermophysical properties, taken at 300 K (See Table \ref{tab:properties}).
\end{itemize}
\begin{table}[!ht]
\begin{center}
\caption{Physical and thermal properties at 300 K.\label{tab:properties}}
    \begin{tabular}{lcccc}
    \hline
    Property                              & \multicolumn{1}{l}{Symbol} & \multicolumn{1}{c}{PDMS} & \multicolumn{1}{c}{Water} & \multicolumn{1}{c}{Copper} \\ \hline
    Density {[}$\text{kg}/\text{m}^3${]} & $\rho$                     & $965$                      & $996.09$                      & $8960$                       \\
    Thermal conductiviy {[}W/mK{]}       & $\kappa$                   & $0.2$                      & $0.61$                     & $401$                        \\
    Specefic heat capacity{[}J/kgK{]}    & $c_p$                      & $1600$                     & $4126.71$                      & $384$                        \\
    Dynamic viscosity{[}Pa.s{]}          & $\mu$                      & -                        & $8.56\times 10^{-4}$                    & -                          \\ \hline
    \end{tabular}
\end{center}    
\end{table}

\item Model 2
\begin{itemize}
    \item Convective and radiative heat transfer from the external PDMS walls to the ambient are neglected.
    \item Compressible fluid, with  thermophysical properties depending on temperature.
\end{itemize}
The dependence of $\rho$, the viscosity $\mu$, the specific heat capacity $c_p$ and the thermal conductivity $\kappa_{f}$ on temperature are obtained from the following empirical functions using data from the literature \cite{the_engineering_toolbox_2004_water_mu,the_engineering_toolbox_2004_water_rho,the_engineering_toolbox_2004_water_kappa,the_engineering_toolbox_2004_water_cp}. 
\begin{equation}\label{eq:rho}
    \rho(T) = 749.09\frac{\text{kg}}{\text{m}^3} + 1.907\frac{\text{kg}}{\text{m}^3\text{K}}T -3.611\times 10^{-3}\frac{\text{kg}}{\text{m}^3\text{K}^2}T^2
\end{equation}
\begin{equation}\label{eq:mu}
\begin{aligned}
    \mu(T) = {} & 1.128 \times 10^{-1}\text{Pa\,s} - 9.628\times 10^{-4}\frac{\text{Pa\,s}}{\text{K}}T + \\
                & 2.758\times10^{-6}\frac{\text{Pa\,s}}{\text{K}^2}T^2 - 2.644\times10^{-9}\frac{\text{Pa\,s}}{\text{K}^3}T^3    
\end{aligned}
\end{equation}
\begin{equation}\label{eq:kappa}
    \kappa(T) = -7.522\times 10^{-1}\frac{\text{W}}{\text{m\,K}} +7.437\times10^{-3}\frac{\text{W}}{\text{m\,K}^2}T -9.674\times10^{-6}\frac{\text{W}}{\text{m\,K}^3}T^2
\end{equation}
\begin{equation}\label{eq:cp}
    c_p(T) = 4099.09\frac{\text{J}}{\text{kg\,K}} -4.166\frac{\text{J}}{\text{kg\,K}^2}T + 1.358\times 10^{-2}\frac{\text{J}}{\text{kg\,K}^3}T^2
\end{equation}
In Fig.\ref{fig:thermophysical} $\rho$, $\mu$,  $c_p$ and $\kappa_{f}$ are plotted as a function of the temperature according to Eqs. \ref{eq:rho}, \ref{eq:mu}, \ref{eq:kappa} and \ref{eq:cp}.
\begin{figure}[H]
    \centering
    \includegraphics[scale=0.4]{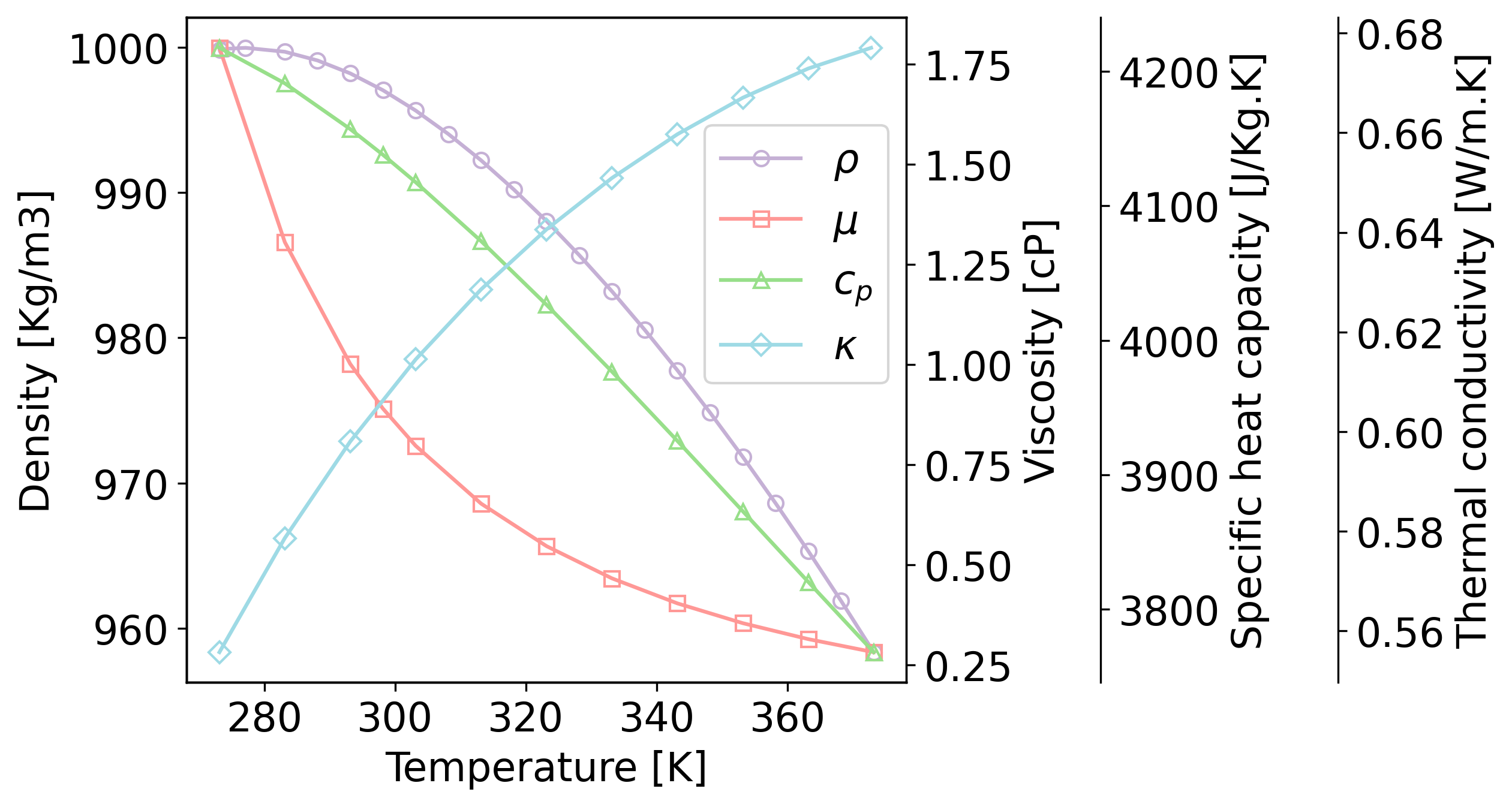}
    \caption{Water thermophysical properties}
    \label{fig:thermophysical}
\end{figure}

\item Model 3
\begin{itemize}
    \item Convective heat transfer from the external PDMS walls to the ambient is considered, but radiative transfer is neglected.
    \item Compressible fluid, with thermophysical properties depending on temperature (given by Eqs. \ref{eq:rho}, \ref{eq:mu}, \ref{eq:kappa} and \ref{eq:cp}.).
\end{itemize}
In this model, the heat can be transferred from the external walls to the environment, so the system is no longer adiabatic. The heat flux due to convection (Newton's cooling law \cite{newton1701scala}) is applied:
\begin{equation}
    \vec{q}_{\text{conv}} = h_{\text{conv}}(T_{\text{walls}}-T_{\text{amb}} )
\end{equation}
Here $h_{\text{conv}}$ is the typical heat transfer coefficient for free air convection, with $h_{\text{conv}} = 10\, \text{W/m}^2\text{K}$, $T_{\text{amb}} = 300\, \text{K}$ representing the temperature of the surrounding air and $T_{\text{walls}}$ is the temperature at the PDMS walls, which can vary depending on the proximity to the hotspot.

\item Model 4
\begin{itemize}
    \item Convective and radiative heat transfer from the external PDMS walls to the ambient are considered.
    \item Compressible fluid with  thermophysical properties depending on temperature (given by Eqs. \ref{eq:rho}, \ref{eq:mu}, \ref{eq:kappa} and \ref{eq:cp}.)
\end{itemize}

In addition to the free convection by air, this model considers a radiative heat transfer from the outer walls of the device.  Using the Stefan–Boltzmann law of radiation \cite{boltzmann1877beziehung} the heat flux from the PDMS surface to the ambient can be expressed as:
\begin{equation}
    \vec{q}_{\text{rad}} = \epsilon \sigma (T^4_{\text{sup}} - T^4_{\text{amb}} ) \ \hat{n}
\end{equation}
where $\epsilon$ is the surface emissivity, $\sigma= 5.67 \times 10 ^{-8}$ W/(m$^2$ K$^4$) is the Stephan-Boltzmann constant, and $\hat{n}$ is the normal unit vector.  Assuming that all points of the surface have temperatures slightly above the ambient temperature, we have $T_{\text{walls}} = T_{\text{amb}} + \Delta T$, with $\Delta T \ll T_{\text{amb}}$, therefore we can perform an expansion up to second order:
\begin{equation}
\vec{q}_{\text{rad}} = \epsilon \sigma \Big( 4 \ T_{\text{amb}}^3 \Delta T + 6 \  T_{\text{amb}}^2 ( \Delta T)^2 + \cdots \Big) \ \hat{n}
\end{equation}
The first term of this equation, linear in $\Delta T$, is analogous to the convective Newton's cooling law. Therefore we can define a radiative coefficient as:
\begin{equation}
    h_{\text{rad}} = 4 \epsilon \sigma  T_{\text{amb}}^3
\end{equation}

Given that PDMS presents emissivities in the range $0.75-0.95$  for wavelengths going from $8-13$ \SI{}{\micro\metre} (main atmospheric window) \cite{tu_antireflection_2023}, we consider $\epsilon=0.85$. Then, for an ambient temperature of 300K, we obtain a radiative coefficient $h_{\text{rad}}$ = 5W/(m$^2$K).

Adding the convective and radiative contributions, we can define an effective heat transfer coefficient $h_{\text{tot}}=15\, \text{W/m}^2\text{K}$ \cite{kreith2001principles}.  Therefore, the total heat flux can be expressed as:
\begin{gather}
    \vec{q}_{\text{total}} = \vec{q}_{\text{conv}} + \vec{q}_{\text{rad}}\\
    \vec{q}_{\text{total}} = (h_{\text{conv}} + h_{\text{rad}})(T_{\text{walls}}-T_{\text{amb}})\\
    \vec{q}_{\text{total}} = h_{\text{tot}}(T_{\text{walls}}-T_{\text{amb}} )
    \label{eq:heff}
\end{gather}
\end{itemize}

\subsection{Solution methodology}\label{sec:solutionMethodology}
\begin{figure}[H]
     \centering
     \begin{subfigure}[b]{0.495\textwidth}
         \centering
         \includegraphics[width=\textwidth]{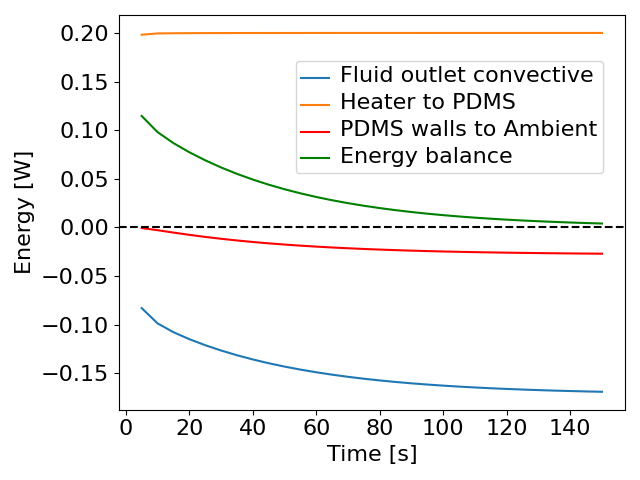}
         \caption{Energy balance Model $3$, $\Delta \text{P}=1000$Pa }
         \label{fig:balance}
     \end{subfigure}
     \hfill
     \begin{subfigure}[b]{0.495\textwidth}
         \centering
         \includegraphics[width=\textwidth]{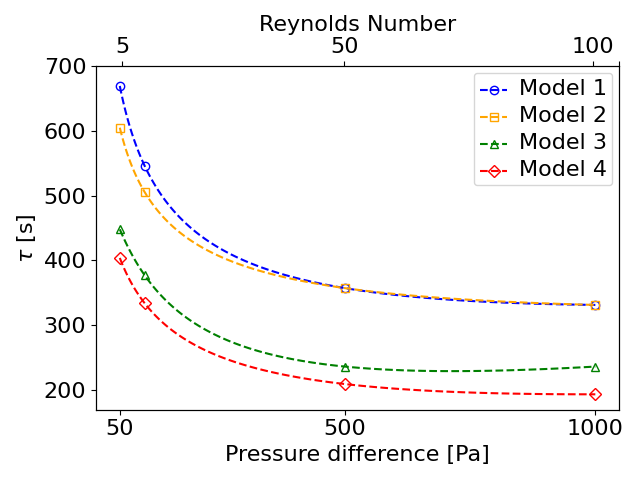}
         \caption{Stady state reachest time for all models}
         \label{fig:tau}
     \end{subfigure}
        \caption{Steady state time}
        \label{fig:steadyStateTime}
\end{figure}

The study of the heat transfer problem requires to solve simultaneously all the relevant solid and flow field heat transfer processes given by Eqs.  \ref{eq:mass}, \ref{eq:momentum}, \ref{eq:energy} jointly with the corresponding boundary conditions. 
The numerical solution of the computational fluid dynamics (CFD) and the conjugate heat transfer problem was performed using the Finite Volume Method (FVM) with the open-source software OpenFOAM  \cite{openfoam}.
In order to get a more realistic approach, simulations are performed considering both the transient and the stationary regime. 
The temporal accuracy was achieved through the use of a second-order temporal scheme integrated with the looping of a PIMPLE algorithm \cite{greenshieldsweller2022}, which facilitated the precise discretization of the momentum and energy equations. The convergence criteria for the stationary regime was based on the energy balance between the power dissipated by the hotspot and the energy dissipated by the device. Additionally, the criterion for the computational convergence was checked with velocity and energy final residuals less than $10^{-7}$.
The steady state criterion must be checked simultaneously in all the regions of the device, in particular when considering materials with a low thermal conductivity as PDMS. As an example, in Fig.\ref{fig:balance} we show the energy calculation for Model 3 for $\Delta \text{P}=100$ Pa. In this case, the heat dissipated through water plus the heat dissipated through the walls by convection with the air, should balance the heat from the hotspot to the PDMS device. The balance is achieved for times larger than a critical time $\tau$ that depends on the parameters of the model and it is found as: 
\begin{equation}
    \frac{|T_{\text{out}}(t)-T_{\text{out}}(t+\Delta t)|}{T_{\text{out}}(t)} \leq 0.001  \,\,\,\,  \,\,\,\,  \forall \,\,\,t > \tau
\end{equation}
Temperature is chosen as the magnitude for the convergence criterion due to its larger time to reach the steady state when compare with other magnitudes such as energy or velocity. This rigorous procedure is mandatory to be repeated and checked systematically for all simulations.
In Fig.\ref{fig:tau} $\tau$ versus $\Delta \text{P}$ is plotted for all the models. It is observed that $\tau$ presents a monotonically decrease in all cases, with lower values as long the model incorporates more realistic assumptions (Model $4$).
The flow is produced by a pressure difference between the inlet and outlet of the microchannel. For simulations, $\Delta \text{P}$ ranges from $\Delta \text{P}=50\, \text{Pa}$ to $\Delta \text{P}=1000$ along the 50 mm long microchannel, values usually found in the literature of thermal microfluidics.
The fluid temperature at the inlet of the microchannel is kept constant at $300$K for all the models and for all times. At time $t = 0$, all parts of the system (PDMS, copper hotspot, fluid) are at  thermal equilibrium with the environment at $T_{\text{amb}}=300$ K. For $t>0$  temperature field evolves according to the model considered. 
Additionally, it is assumed that initially the fluid is at rest, so $\Vec{u}=(0,0,0)\frac{\text{m}}{\text{s}}$.

\begin{figure}[H]
    \centering
    \includegraphics[scale=0.2]{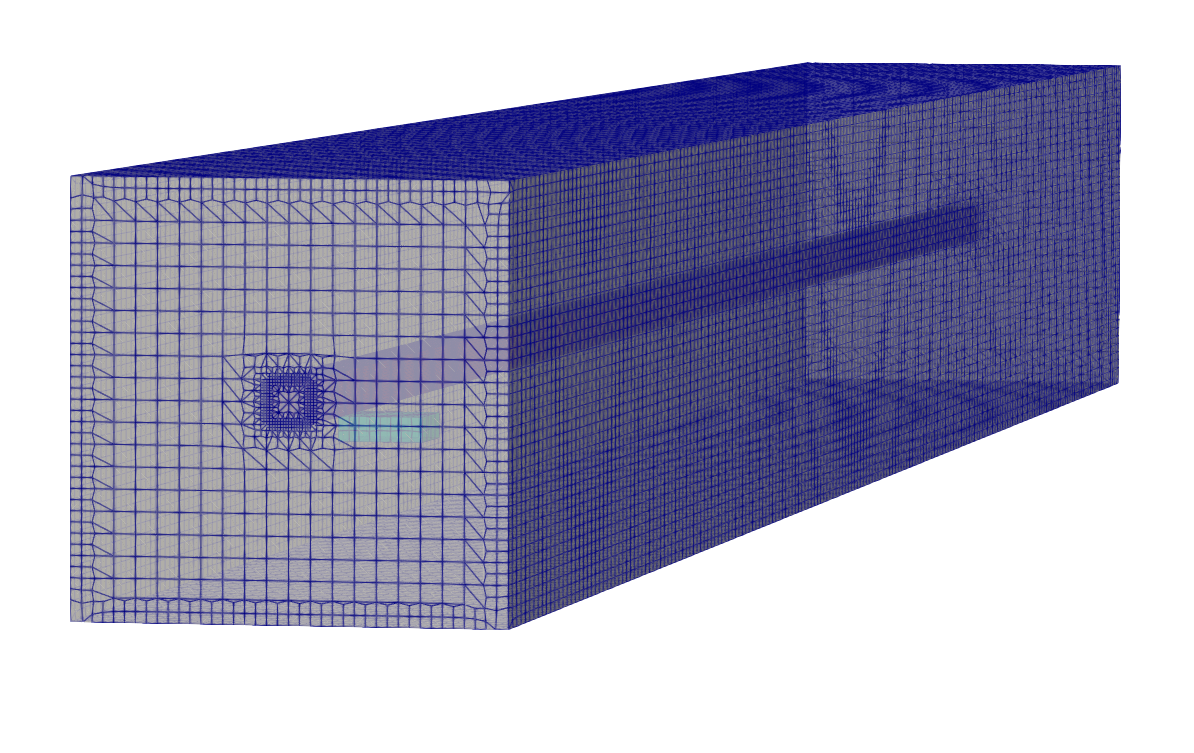}
    \caption{Computational mesh for simulation domain of the microfluidic device}
    \label{fig:mesh}
\end{figure}

For the simulations we consider a $3\text{D}$ mesh model (see Fig. \ref{fig:mesh}) that was generated using an OpenFoam meshing function to optimize computational efficiency. The refinement of the mesh at the surface regions interfaces is crucial to capture with good accuracy the underlying phenomena\cite{forsbergChapterForcedConvection2021}.

As it was mentioned above, the heat transfer from the PDMS to the ambient, can be calculated using a convection boundary condition with an effective heat transfer coefficient given by Eq.\ref{eq:heff}. 
From a computational point of view this effective approach is extremely advantageous, because it is not necessary to simulate a large volume of air surrounding the device, which would entail a significant computational cost.

\section{Results}\label{sec:results}
\begin{figure}[H]
    \centering
    \includegraphics[scale=0.45]{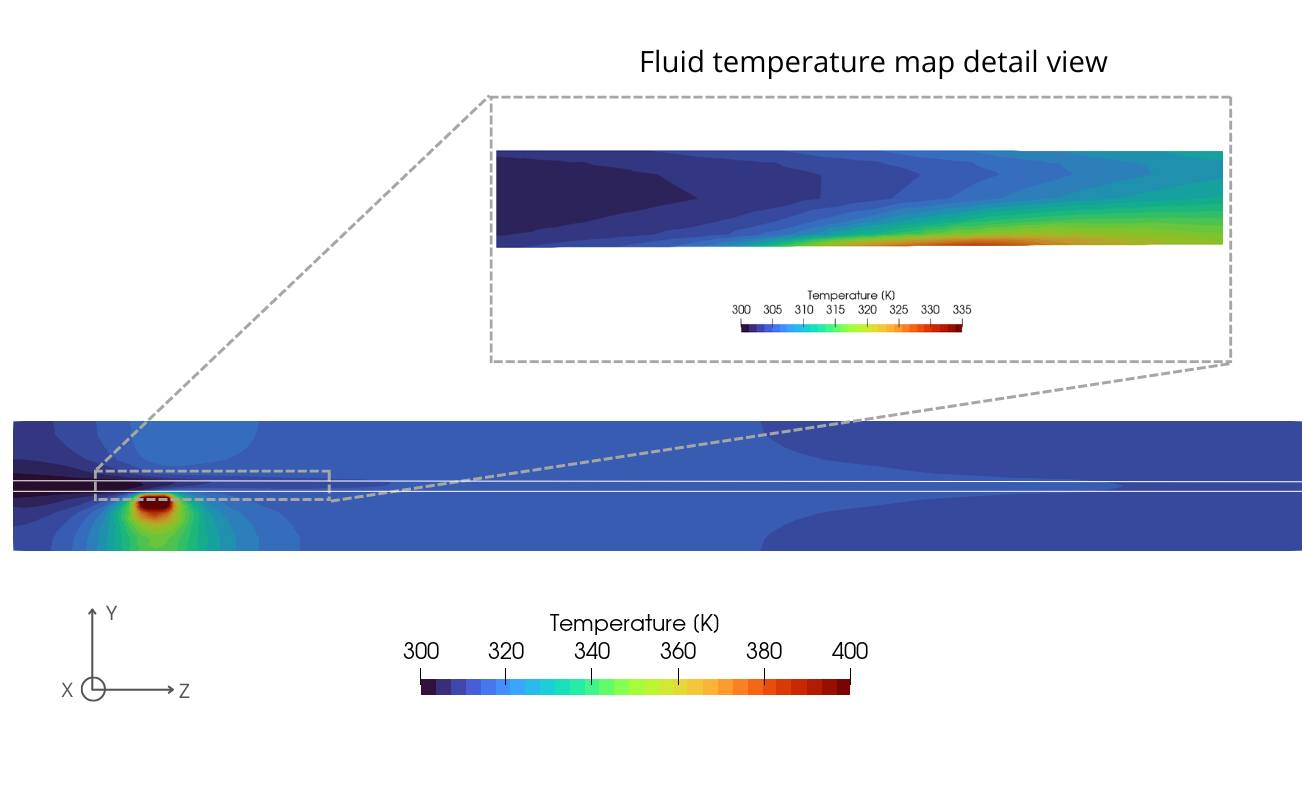}
    \caption{Cross section temperature map for PDMS-microfluid device and fluid detail view (Model $4$). Red zone corresponds to the location of the hotspot.}
    \label{fig:temperature}
\end{figure}

We analyze the heat transfer and energy balance for the four different models described in Section \ref{sec:physicalModel} as a function of the pressure difference. Fig.\ref{fig:temperature} shows the temperature map on a cross-section along the whole device traversing the copper hotspot, the water channel and the PDMS, for $\Delta \text{P}=1000 \text{Pa}$. The inset corresponds to a detail of the fluid at region near the hotspot. As the temperature maps corresponding to the different models do not present qualitative differences, we choose as demonstrative model the one with most of the contributions (Model $4$).

\begin{figure}[H]
     \centering
     \begin{subfigure}[b]{0.475\textwidth}
         \centering
         \includegraphics[width=\textwidth]{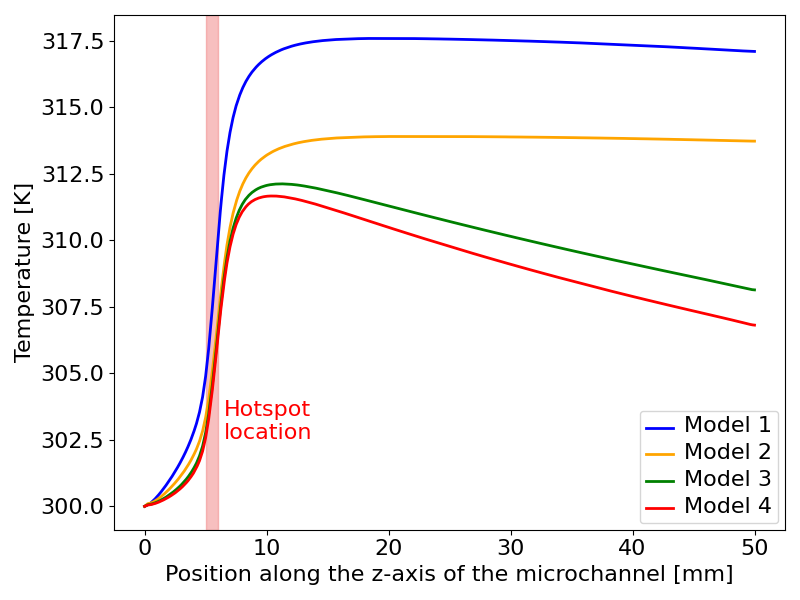}
         \caption{All models at $\Delta \text{P}=50$Pa}
         \label{fig:Tvsl_models}
     \end{subfigure}
     \hfill
     \begin{subfigure}[b]{0.475\textwidth}
         \centering
         \includegraphics[width=\textwidth]{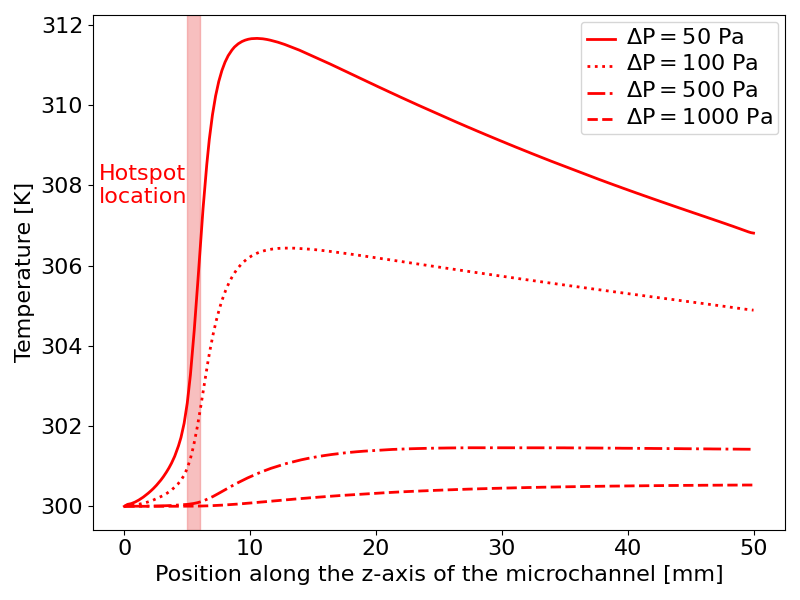}
         \caption{Model $4$ at different $\Delta \text{P}$}
         \label{fig:Tvsl_pressures}
     \end{subfigure}
        \caption{Fluid temperature along the channel (z-axis), measured at the center of the microchannel. Red shaded zone corresponds to the location of the hotspot.}
        \label{fig:Tvsl}
\end{figure}

Thermophysical properties of the fluid such as density and viscosity, and consequently the velocity profiles, are affected directly by the temperature changes. This effect is enhanced in the area near the hotspot, as can be seen in Fig. \ref{fig:Tvsl}.

As viscosity decreases with temperature the fluid moves faster, increasing the flow rate. This is the case for the Model 2 in which temperature is lower than in Model 1, as can be seen in Fig. \ref{fig:Tvsl_models}.
The heat transfer through the fluid is reduced if losses through the walls due to convection/radiation are taken into account. This effect is strongly dependent on Re ($\Delta \text{P}$), as seen in Fig. \ref{fig:Tvsl_pressures}. It is observed that an increase of $\Delta \text{P}$ results in a lower temperature along the channel. However the dependence on the position $z$ along the microchannel displays different qualitatively behaviours. For low Re the temperature presents a non monotonic behaviour. For higher Re numbers the response is monotonic. In this case the temperature grows near the hotspot and saturates reaching a constant value along the channel. It is interesting to note that at very low Re in all models, upstream heating occurs. Due to the very low velocity, heat diffusion becomes significant compared to the other heat transfer mechanisms.

In steady state, the different parts of the system acquire a maximum temperature (after a transient) given by the balance between the power injected and the heat removed through the different transfer mechanisms. In particular, with respect to the operational regime of the device as a heat exchanger, it is important to analyze the dependence of the maximum temperature at the hotspot and at the fluid outlet on the external pressure difference under which the device operates.
In several experimental works \cite{qu_experimental_2002,khoshvaght-aliabadi_experimental_2016,huang_experimental_2020} "chambers" or "plenums" are often used in front of and behind the microchannels, fulfilling different functions. One of them is to mix the fluid correctly both upstream and downstream to measure its temperature. In this way, thermocouples are usually placed in these basins to have a uniform measurement of the inlet and outlet temperatures of the fluid. However if the temperature is not uniform, and in order to possible comparisons with experiment it is required the calculation of an average value obtained by integration of the local temperature over the outlet cross section using the following expression:  

\begin{equation}\label{eq:outletTemperature}
        T_{\text{out}} = \frac{\int_A \rho_{\text{cell}}v^{\text{cell}}_z T_{\text{cell}} dA}{\int_A \rho_{\text{cell}}v^{\text{cell}}_z dA}
\end{equation}
where the "cell" suffixes indicate the value of the property (temperature, velocity, density) evaluated at a specific unit cell at the mesh in the FVM simulation. If the outlet is far from the hotspot, in general the temperature profile is practically homogeneous. 

Accordingly the mean heat flow evacuated by the fluid at the outlet as a result of convection can be calculated as:
\begin{gather}
q_{\text{out}}=\sum_{\text{cell}=1}^nq_{\text{cell}}\\
    q_{\text{cell}}=A_{\text{cell}}c_p(T)v^{\text{cell}}_z\rho(T)(T_{\text{cell}}-T_{\text{in}})\\
    T_{\text{cell}} = T_{\text{in}} + \frac{q_{\text{cell}}}{A_{\text{cell}}c_{p_{\text{cell}}}(T)v^{\text{cell}}_z\rho_{\text{cell}}(T)}\label{eq:tOut}
\end{gather}

\begin{figure}[H]
     \centering
     \begin{subfigure}[b]{0.475\textwidth}
         \centering
         \includegraphics[width=\textwidth]{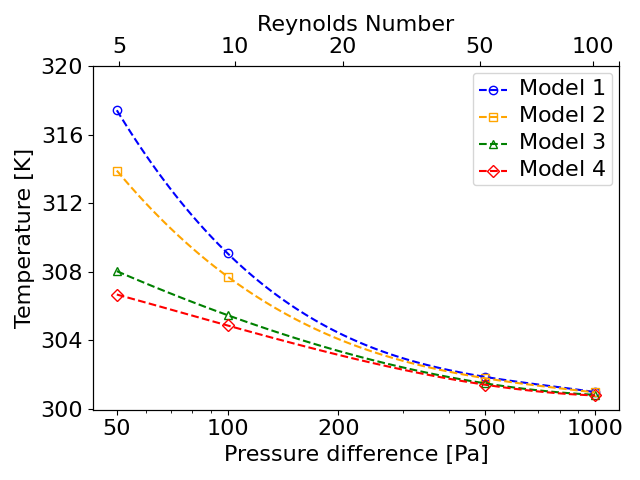}
         \caption{Outlet fluid temperature}
         \label{fig:outTempModels}
     \end{subfigure}
     \hfill
     \begin{subfigure}[b]{0.475\textwidth}
         \centering
         \includegraphics[width=\textwidth]{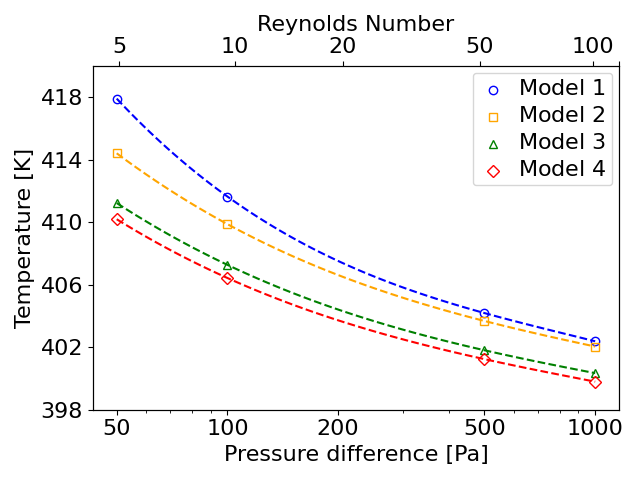}
         \caption{Maximum copper temperature}
         \label{fig:maxTempModels}
     \end{subfigure}
        \caption{Temperature reached by the coolant at the microchannel outlet Eq.(\ref{eq:outletTemperature} $a)$ and by the hotspot $b)$, when reaching the time $\tau$ for all models and different $\Delta P$.}
        \label{fig:modelsComparasion}
\end{figure}
\begin{figure}[H]
     \centering
     \begin{subfigure}[b]{0.495\textwidth}
         \centering
         \includegraphics[width=\textwidth]{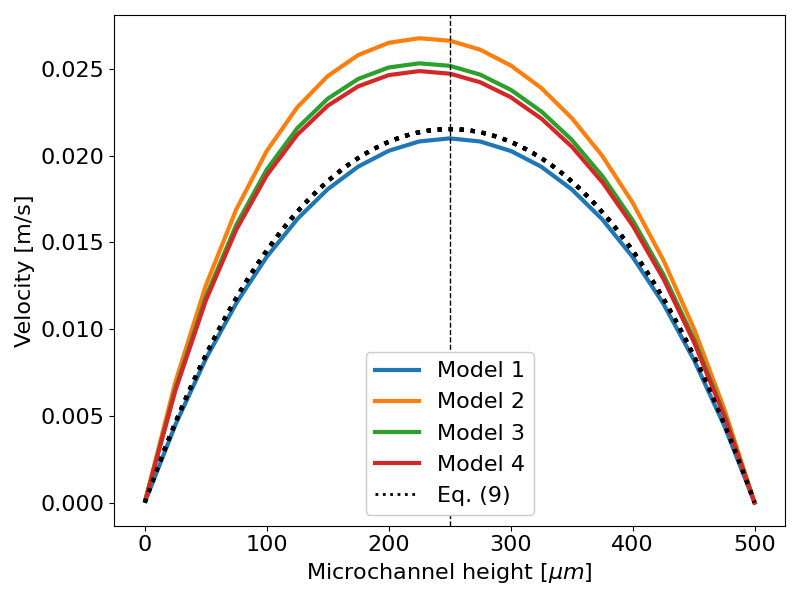}
         \caption{hotspot region}
         \label{fig:hotspotVelocities}
     \end{subfigure}
     \hfill
     \begin{subfigure}[b]{0.495\textwidth}
         \centering
         \includegraphics[width=\textwidth]{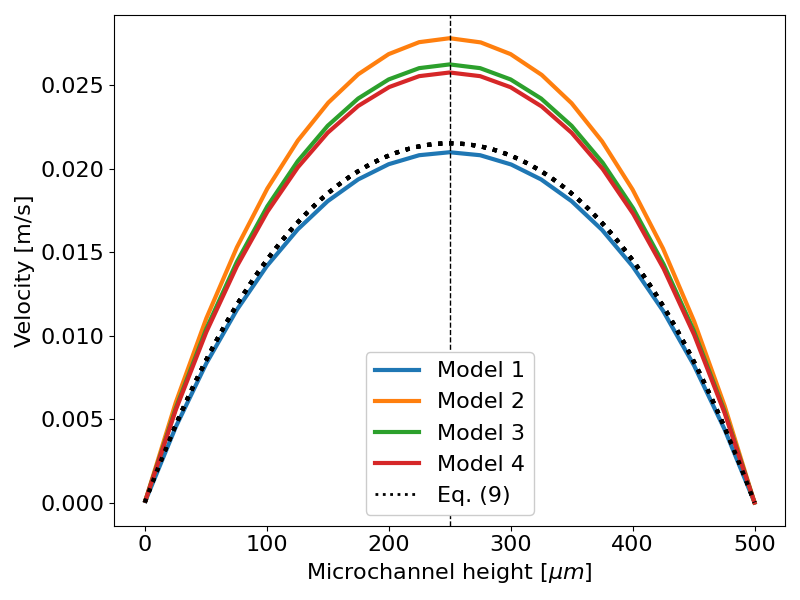}
         \caption{outlet region}
         \label{fig:outletVelocities}
     \end{subfigure}
        \caption{Poiseuille velocity profile for all models at $\Delta \text{P}=50 \text{Pa}$}
        \label{fig:poiseuille}
\end{figure}

Taking into account these considerations, we calculate the maximum temperatures at the outlet and at the hotspot as a function of the pressure difference $\Delta \text{P}$ for all the models. The results are shown in Fig. \ref{fig:modelsComparasion}. 
In both figures, the temperature displays a monotonically decrease with Re. When $\Delta \text{P}$ increases the curves corresponding to the four models approach each other, as long at the temperature tends to the ambient value (Fig. \ref{fig:outTempModels}).  For small Re, temperature is lower at the outlet as well as at the hotspot (Fig. \ref{fig:maxTempModels}), as an indication that radiative and free air convection become more relevant mechanism for heat transfer.
In connection to temperature, it is also interesting to analyze the velocity profiles at the hotspot and outlet positions.  In Fig.\ref{fig:hotspotVelocities} the velocity is calculated at a cross section located at hotspot position and at the outlet region Fig.\ref{fig:outletVelocities}.

\begin{figure}[H]
    \centering
    \includegraphics[scale=0.5]{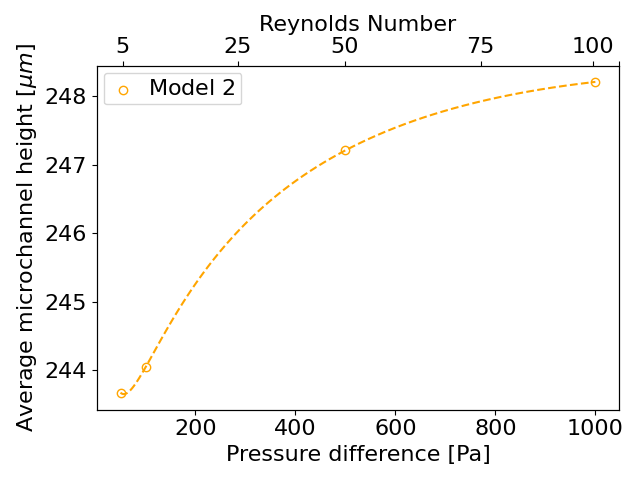}
    \caption{Average height (Eq. \ref{eq:h_med}) for Model 2 at different Reynolds numbers as a measure of the velocity profile asymmetry}
    \label{fig:assymetry}
\end{figure}

In Fig. \ref{fig:poiseuille} the black dotted line correspond to the theoretical Poiseuille velocity profile given by Eq.\ref{eq:rectangular}. It can be observed that Model $1$ is the more similar to the theoretical curve, with a good agreement both in values and shape. The other three models, that consider more realistic assumptions, present a more significative difference respect to the theoretical curve. Moreover, near the hotspot these three models are not symmetric about the $y-$axis. The considerable variations of the temperature at this region produce significant changes in the local thermophysical properties, which affects the symmetry of the velocity profile (see Fig.\ref{fig:hotspotVelocities}). However, this asymmetry is washed up as the distance to the hotspot increases. 
In Fig. \ref{fig:assymetry} it is observed that for Model 2 (chosen because it presents the most notorious deviations in Fig \ref{fig:hotspotVelocities}), as the pressure difference decreases, the average microchannel height of the velocity profile at the area near the hotspot turns away from the middle of the channel height ($250 \mu\text{m}$).

If the profile were symmetric, the maximum velocity should be at half of the microchannel height. As the profile was found not to be symmetric, we calculated a weighted height with the velocity profile in order to quantify the asymmetry of the profile:

\begin{equation}
\Bar{h}= \frac{\sum_{i}^{n} v_z(h_{i})h_i}{\sum_{i}^{n}v_z(h_{i})}
\label{eq:h_med}
\end{equation}

where $v_z(h_i)$ is the z-axis component of the velocity as a function of microchannel height, ranging from 0 to 500 $\mu m$.

The mass flow rate at the outlet can also be calculated for each model as: 
\begin{equation}\label{eq:massFlowRate}
    \dot{m} = \sum_{\text{cell}=1}^{n} A_{\text{cell}}\rho(T_{\text{cell}})v^{\text{cell}}_z
\end{equation}

It is found that the mass flow rate presents a linear response with the pressure difference for all the models. However, in the very low Re regime the computed values show large differences for the different models. In this case, heat removal at the outlet by convection is no longer the main mechanism of heat transfer, becoming radiation and wall air convection more relevant. Taking Model $1$ as a reference with $\Delta \text{P}=50\text{Pa}$, the mass flow rate for Model $2$ has a variation of $31.73\%$, Model $3$ $23.30\%$ and Model $4$ $21.28\%$. This is consistent with results in Fig. \ref{fig:poiseuille}, due to the linear dependence of the mass flow rate on the velocity, shown in Eq. \ref{eq:massFlowRate}. Moreover, the agreement between the mass flow rate obtained from Model $1$ and Eq. \ref{eq:caudal} was checked, finding a maximum deviation of $3.82\%$ for $\Delta \text{P}=50 \text{Pa}$.
In contrast to Figs. \ref{fig:outTempModels} and \ref{fig:maxTempModels} in which the highest temperature  corresponds to the simplest model (Model $1$), in Fig. \ref{fig:poiseuille} it is observed just the opposite for the velocity of this model.  This is because the thermophysical properties of the fluid for that model remain constant (see Table \ref{tab:properties}) evaluated at the inlet temperature ($300$K). On the other hand, as velocity is inversely proportional to the viscosity (Eq. \ref{eq:rectangular}), larger velocities and flow rates are obtained.

\begin{figure}[H]
    \centering
    \includegraphics[scale=0.5]{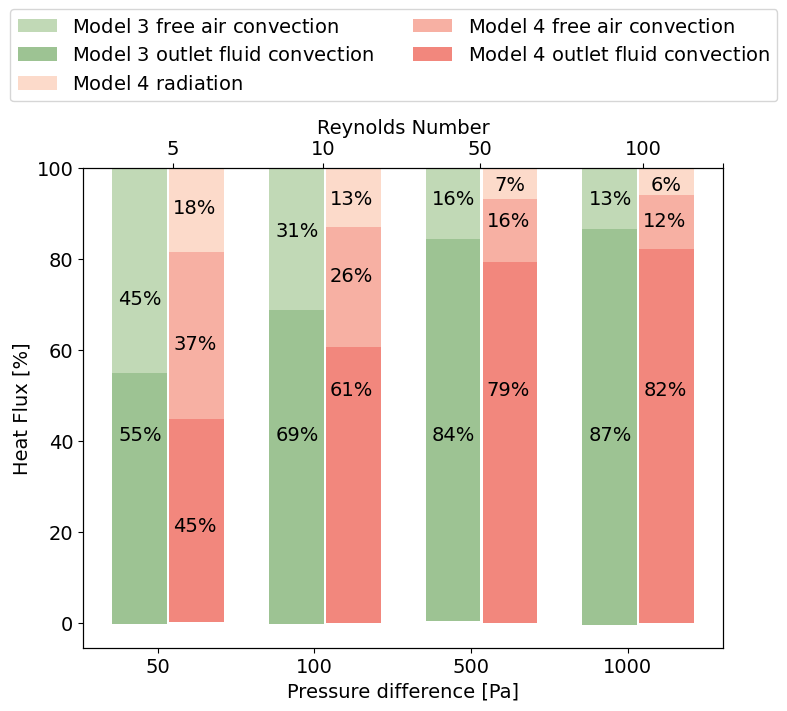}
    \caption{Heat transfer mechanism percent contribution for Model $3$ and Model $4$}
    \label{fig:heatFluxBalance}
\end{figure}

The results evidence that depending on the Re regime considered, the different models can yield significant differences.  To analyze this point we focus on the heat transport contribution due to the radiative and convective components. Model $1$ and $2$ are not considered in this analysis because as the device is isolated, the fluid outlet convection is the only mechanism to dissipate heat. 
In Fig.\ref{fig:heatFluxBalance} it is compared the percent contribution to the heat fluxes for Models $3$ and $4$. 
For both models, the convective and radiative contributions are significant, in particular for very small Re. As the material plays a key role on this contribution,  the modeling has to include the material in order to make accurately comparison with experimental data. For larger Re number, although both contributions are smaller, they are still a considerable fraction percent of the outlet fluid estimations. 

\section{Concluding Remarks}\label{sec:concludingRemarks}

Four models of a PDMS-thermal microfludic setup were proposed to study numerically the effect of different transfer mechanisms and thermophysical properties. It was performed a critical analysis of the assumptions and conventions usually found in the literature when modeling heat transfer along thermal setups consisting of a single straight  microchannel. A detailed discussion was presented about conditions where they fail to capture the full complexity of the problem.

The simpler model was an adiabatic system with constant thermophysical properties. This is the more frequent description found in the literature, even for modeling the overall thermal properties of an entire device. As a second step towards a more realistic model, temperature-dependent thermophysical properties were considered. Finally two models including free air convection at the boundaries of the surrounding material, temperature-dependent properties and thermal radiation were studied.

It was performed a comparative analysis of the
magnitudes required to characterize the thermal performance of a thermal exchanger such as temperature, velocity and flow mass as a function of the applied pressure. The study focused on two specific positions: at the hotspot, location of the power injected, and at the outlet of the fluid channel where heat is removed by convection.

\begin{itemize}
    \item For small Re numbers, the differences between adiabatic and non-adiabatic models are considerably more pronounced. Convection and convection plus radiation result in a reduction of the fluid outlet temperature. Consequently, the loses due to these contributions should be taken into account in order to make an accurate description in single microchannels setups.
    \item Considering the more realistic models, for Re values less than $100$, it is observed a deviation from the Poiseuille velocity profile nearby the hotspot region.
    \item As shown in Figures \ref{fig:modelsComparasion}, \ref{fig:assymetry} and \ref{fig:heatFluxBalance}, the trend indicates that for Re $> 100$, simpler models can be consider accurately enough.
\end{itemize}

In summary, to neglect thermal transfer mechanisms to the environment facilitates calculations and simulations. However, this approach must be done based on a careful qualitative and quantitative analysis considering the characteristics of the system as a whole: fluid thermophysical properties, geometry, materials and environment. The models studied here, although simple, show that this critical analysis is mandatory to make reliable estimations of the overall thermal transfer performance of thermal setups.

\section*{\textbf{Acknowledgments}}
The authors thank to CSC-CONICET to facilitate the cluster TUPAC to perform the simulations.

\bibliographystyle{unsrt}  

\bibliography{sn-bibliography}

\begin{thebibliography}{10}

\bibitem{barakoIntegratedNanomaterialsExtreme2018}
Michael~T. Barako, Vincent Gambin, and Jesse Tice.
\newblock Integrated nanomaterials for extreme thermal management: a perspective for aerospace applications.
\newblock {\em Nanotechnology}, 29(15):154003, February 2018.
\newblock Publisher: IOP Publishing.

\bibitem{yeExperimentalInvestigationsThermal2022}
Yuxin Ye, Binbin Jiao, Yanmei Kong, Ruiwen Liu, Xiangbin Du, Kunpeng Jia, Shichang Yun, and Dapeng Chen.
\newblock Experimental investigations on the thermal superposition effect of multiple hotspots for embedded microfluidic cooling.
\newblock {\em Applied Thermal Engineering}, 202:117849, February 2022.

\bibitem{sohelmurshedCriticalReviewTraditional2017}
S.~M. Sohel~Murshed and C.~A. Nieto~de Castro.
\newblock A critical review of traditional and emerging techniques and fluids for electronics cooling.
\newblock {\em Renewable and Sustainable Energy Reviews}, 78:821--833, October 2017.

\bibitem{lorenzini_embedded_2016}
Daniel Lorenzini, Craig Green, Thomas~E. Sarvey, Xuchen Zhang, Yuanchen Hu, Andrei~G. Fedorov, Muhannad~S. Bakir, and Yogendra Joshi.
\newblock Embedded single phase microfluidic thermal management for non-uniform heating and hotspots using microgaps with variable pin fin clustering.
\newblock 103:1359--1370.

\bibitem{lee_enhanced_2012}
Y.~J. Lee, P.~S. Lee, and S.~K. Chou.
\newblock Enhanced thermal transport in microchannel using oblique fins.
\newblock 134(10).

\bibitem{khoshvaght-aliabadi_experimental_2016}
M.~Khoshvaght-Aliabadi, M.~Sahamiyan, M.~Hesampour, and O.~Sartipzadeh.
\newblock Experimental study on cooling performance of sinusoidal–wavy minichannel heat sink.
\newblock 92:50--61.

\bibitem{grafner_flow_2022}
S.~J. Gräfner, P.~Y. Wu, and C.~R. Kao.
\newblock Flow in a microchannel filled with arrays of numerous pillars.
\newblock 97:109045.

\bibitem{carvalho_computational_2021}
Violeta Carvalho, Raquel~O. Rodrigues, Rui~A. Lima, and Senhorinha Teixeira.
\newblock Computational simulations in advanced microfluidic devices: A review.
\newblock 12(10):1149.

\bibitem{gao_fluid_2022}
Jie Gao, Zhuohuan Hu, Qiguo Yang, Xing Liang, and Hongwei Wu.
\newblock Fluid flow and heat transfer in microchannel heat sinks: Modelling review and recent progress.
\newblock 29:101203.

\bibitem{hung_optimal_2012}
Tu-Chieh Hung, Tsung-Sheng Sheu, and Wei-Mon Yan.
\newblock Optimal thermal design of microchannel heat sinks with different geometric configurations.
\newblock 39(10):1572--1577.

\bibitem{sidik_overview_2017}
Nor Azwadi~Che Sidik, Muhammad Noor Afiq~Witri Muhamad, Wan Mohd Arif~Aziz Japar, and Zainudin~A. Rasid.
\newblock An overview of passive techniques for heat transfer augmentation in microchannel heat sink.
\newblock 88:74--83.

\bibitem{kumar_numerical_2019}
Pankaj Kumar.
\newblock Numerical investigation of fluid flow and heat transfer in trapezoidal microchannel with groove structure.
\newblock 136:33--43.

\bibitem{raghuraman_influence_2017}
D.~R.~S. Raghuraman, R.~Thundil Karuppa~Raj, P.~K. Nagarajan, and B.~V.~A. Rao.
\newblock Influence of aspect ratio on the thermal performance of rectangular shaped micro channel heat sink using {CFD} code.
\newblock 56(1):43--54.

\bibitem{jungExperimentalInvestigationHeat2021}
Sung~Yong Jung and Hanwook Park.
\newblock Experimental investigation of heat transfer of {{Al2O3}} nanofluid in a microchannel heat sink.
\newblock {\em International Journal of Heat and Mass Transfer}, 179:121729, November 2021.

\bibitem{jungHeatTransferFlow2019}
Sung~Yong Jung, Jun~Hong Park, Sang~Joon Lee, and Hanwook Park.
\newblock Heat transfer and flow characteristics of forced convection in {{PDMS}} microchannel heat sink.
\newblock {\em Experimental Thermal and Fluid Science}, 109:109904, December 2019.

\bibitem{oudahExperimentalInvestigationEffect2020}
Saad~K. Oudah, Ruixian Fang, Amitav Tikadar, Azzam~S. Salman, and Jamil~A. Khan.
\newblock An experimental investigation of the effect of multiple inlet restrictors on the heat transfer and pressure drop in a flow boiling microchannel heat sink.
\newblock {\em International Journal of Heat and Mass Transfer}, 153:119582, June 2020.

\bibitem{wangExperimentalNumericalStudy2018}
Yingying Wang, Jeong-Heon Shin, Corey Woodcock, Xiangfei Yu, and Yoav Peles.
\newblock Experimental and numerical study about local heat transfer in a microchannel with a pin fin.
\newblock {\em International Journal of Heat and Mass Transfer}, 121:534--546, June 2018.

\bibitem{yiPDMSNanocompositesHeat2014}
Pyshar Yi, Robiatun~A. Awang, Wayne S.~T. Rowe, Kourosh {Kalantar-zadeh}, and Khashayar Khoshmanesh.
\newblock {{PDMS}} nanocomposites for heat transfer enhancement in microfluidic platforms.
\newblock {\em Lab on a Chip}, 14(17):3419--3426, July 2014.

\bibitem{liEffectsThermalProperty2007}
Zhigang Li, Xiulan Huai, Yujia Tao, and Huanzhuo Chen.
\newblock Effects of thermal property variations on the liquid flow and heat transfer in microchannel heat sinks.
\newblock {\em Applied Thermal Engineering}, 27(17):2803--2814, December 2007.

\bibitem{kumarPhysicalEffectsVariable2018}
Rajan Kumar and Shripad~P. Mahulikar.
\newblock Physical {{Effects}} of {{Variable Thermophysical Fluid Properties}} on {{Flow}} and {{Thermal Development}} in {{Micro-Channel}}.
\newblock {\em Heat Transfer Engineering}, 39(4):374--390, February 2018.

\bibitem{yiPDMSNanocompositesHeat2014a}
Pyshar Yi, Robiatun~A. Awang, Wayne S.~T. Rowe, Kourosh {Kalantar-zadeh}, and Khashayar Khoshmanesh.
\newblock {{PDMS}} nanocomposites for heat transfer enhancement in microfluidic platforms.
\newblock {\em Lab on a Chip}, 14(17):3419--3426, July 2014.

\bibitem{wuPolydimethylsiloxaneMicrofluidicChip2009}
Jinbo Wu, Wenbin Cao, Weijia Wen, Donald~Choy Chang, and Ping Sheng.
\newblock Polydimethylsiloxane microfluidic chip with integrated microheater and thermal sensor.
\newblock {\em Biomicrofluidics}, 3(1):012005, January 2009.

\bibitem{kumarThermalModelingDesign2013}
Sumeet Kumar, Marco~A. {Cartas-Ayala}, and Todd Thorsen.
\newblock Thermal modeling and design analysis of a continuous flow microfluidic chip.
\newblock {\em International Journal of Thermal Sciences}, 67:72--86, May 2013.

\bibitem{linPDMSMicrofabricationDesign2021}
Lin Lin and Chen-Kuei Chung.
\newblock {{PDMS Microfabrication}} and {{Design}} for {{Microfluidics}} and {{Sustainable Energy Application}}: {{Review}}.
\newblock {\em Micromachines}, 12(11):1350, November 2021.

\bibitem{joThreedimensionalMicrochannelFabrication2000}
B.-H. Jo, L.M. Van~Lerberghe, K.M. Motsegood, and D.J. Beebe.
\newblock Three-dimensional micro-channel fabrication in polydimethylsiloxane ({{PDMS}}) elastomer.
\newblock {\em Journal of Microelectromechanical Systems}, 9(1):76--81, March 2000.

\bibitem{qu_analysis_2002}
Weilin Qu and Issam Mudawar.
\newblock Analysis of three-dimensional heat transfer in micro-channel heat sinks.
\newblock 45(19):3973--3985.

\bibitem{ahmedOptimumThermalDesign2015}
Hamdi~E. Ahmed and Mirghani~I. Ahmed.
\newblock Optimum thermal design of triangular, trapezoidal and rectangular grooved microchannel heat sinks.
\newblock {\em International Communications in Heat and Mass Transfer}, 66:47--57, August 2015.

\bibitem{boussinesq1868}
{M. J. Boussinesq}.
\newblock Mémoire sur l'influence des frottements dans les mouvements réguliers des fluides.
\newblock {\em 1868}, 13:377--424.

\bibitem{the_engineering_toolbox_2004_water_mu}
{The Engineering ToolBox (2004)}.
\newblock Water - {Dynamic} ({Absolute}) and {Kinematic} {Viscosity} vs. {Temperature} and {Pressure}.

\bibitem{the_engineering_toolbox_2004_water_rho}
{The Engineering ToolBox (2004)}.
\newblock Water - {Density}, {Specific} {Weight} and {Thermal} {Expansion} {Coefficients}.

\bibitem{the_engineering_toolbox_2004_water_kappa}
{The Engineering ToolBox (2004)}.
\newblock Water - {Thermal} {Conductivity} vs. {Temperature}.

\bibitem{the_engineering_toolbox_2004_water_cp}
{The Engineering ToolBox (2004)}.
\newblock Water - {Specific} {Heat} vs. {Temperature}.

\bibitem{newton1701scala}
I.~Newton.
\newblock {\em Scala graduum caloris: calorum descriptiones \& signa}.
\newblock Royal Society of London.

\bibitem{boltzmann1877beziehung}
L.~Boltzmann.
\newblock {\em Über Die {{Beziehung}} Zwischen Dem Zweiten {{Hauptsatze}} Des Mechanischen {{Wärmetheorie}} Und Der {{Wahrscheinlichkeitsrechnung}}, Respective Den {{Sätzen}} Über Das {{Wärmegleichgewicht}}}.
\newblock {K.k. Hof- und Staatsdruckerei}.

\bibitem{tu_antireflection_2023}
Yiteng Tu, Xinyu Tan, Xiongbo Yang, Guiguang Qi, Kun Yan, and Zhe Kang.
\newblock Antireflection and radiative cooling difunctional coating design for silicon solar cells.
\newblock {\em Optics Express}, 31(14):22296--22307, July 2023.
\newblock Publisher: Optica Publishing Group.

\bibitem{kreith2001principles}
F.~Kreith and M.~Bohn.
\newblock {\em Principles of Heat Transfer}.
\newblock {Brooks/Cole Pub.}

\bibitem{openfoam}
{OpenFOAM} {\textbar} {Free} {CFD} {Software} {\textbar} {The} {OpenFOAM} {Foundation}.

\bibitem{greenshieldsweller2022}
Christopher Greenshields and Henry Weller.
\newblock {\em Notes on Computational Fluid Dynamics: General Principles}.
\newblock CFD Direct Ltd, Reading, UK, 2022.

\bibitem{forsbergChapterForcedConvection2021}
Charles~H. Forsberg.
\newblock Chapter 6 - {Forced} convection.
\newblock In Charles~H. Forsberg, editor, {\em Heat {Transfer} {Principles} and {Applications}}, pages 211--266. Academic Press, January 2021.

\bibitem{qu_experimental_2002}
Weilin Qu and Issam Mudawar.
\newblock Experimental and numerical study of pressure drop and heat transfer in a single-phase micro-channel heat sink.
\newblock 45(12):2549--2565.

\bibitem{huang_experimental_2020}
Binghuan Huang, Haiwang Li, Shuangzhi Xia, and Tiantong Xu.
\newblock Experimental investigation of the flow and heat transfer performance in micro-channel heat exchangers with cavities.
\newblock 159:120075.

\end{thebibliography}

\end{document}